\documentstyle[aps,tighten,epsfig,pstricks]{revtex} 
\begin{document}
\draft
\title{ 
Thermal Parameters in Heavy Ion Collisions
at SPS and RHIC:\\ 
Centrality Dependence} 

\author{
{\sc B.~K\"ampfer$^a$, 
J.~Cleymans$^b$, 
K.~Gallmeister$^{a, c}$,  
S.M.~Wheaton$^b$}}

\address{
$^a$ Forschungszentrum Rossendorf, 
PF 510119, 01314 Dresden, Germany\\
$^b$ University of Cape Town,
Rondebosch 7701, Cape Town, South Africa\\[0.2ex] 
$^c$ 
Institut f\"ur Theoretische Physik, Universit\"at Giessen, Giessen, 
Germany\thanks{present address}
}

\maketitle 

\begin{abstract}
The centrality dependence of thermal parameters
describing hadron multiplicities and intermediate-mass dilepton
spectra in heavy-ion collisions at SPS and RHIC is analyzed.
From experimental hadron multiplicities we deduce evidence for strangeness
saturation at high energy and maximum centrality.
The observed dilepton spectra can be parameterized by a centrality independent
temperature.\\[3mm]
{\it keywords:} relativistic heavy-ion collisions, thermal models,
hadron yields, dileptons\\
{\it PACS:} 24.10.Pa, 25.75.Dw, 25.75.-q, 27.75.+p, 12.38.Mh, 24.85.+p
\end{abstract}
 
\section{Introduction}\label{intro}
It has been shown that various observables 
of relativistic heavy-ion collisions
can be well described by statistical-thermal or hydrodynamical models.
In such a way, a selected subset of a large number of observables
can be reproduced by a small number of characteristic parameters,
such as temperature, density or flow velocity.
It is the subject of the present note to pursue this idea
and to analyze the centrality dependence of the thermal
parameters describing hadron multiplicities and dilepton spectra
in the intermediate-mass region.
This will provide further information 
about the effects of the size of the excited strongly interacting
system and help in the systematic understanding of the experimental data.  

The aim of the present work is twofold: 
(i) the previous analysis \cite{hirschegg_02} of hadron multiplicities at
SPS and RHIC is repeated and improved, e.g., by 
including feed-down from weak decays, and (ii)  
the consideration of the intermediate-mass dilepton spectra at SPS 
is outlined in more detail, in particular by analyzing
both the invariant mass and transverse momentum spectra.
 
\section{Hadron Multiplicities} 
Hadron multiplicities can be reproduced 
\cite{abundance,Becattini} by
the grand-canonical partition function
${\cal Z}_i (V, T, \mu^\alpha) = \mbox{Tr} \left[
\exp\{- (\hat H - \mu^\alpha Q_i^\alpha) / T \} \right]$,
where $\hat H$ is the statistical operator of the system,
$T$ denotes the temperature, and $\mu^\alpha$ and 
$Q_i^\alpha$ represent the 
chemical potentials and corresponding conserved charges
respectively.
In the analysis of $4\pi$ data, the net-zero strangeness and the 
baryon-to-electric charge ratio of the colliding nuclei constrain
the components $\mu^\alpha = (\mu_B, \mu_S, \mu_Q$),
where the subscripts $B, S$ and $Q$ refer to baryon, strangeness, electric
charge. 
These constraints have to be relaxed
when considering data in a limited rapidity window, increasing the
number of free parameters. 
The particle numbers are given by 
\begin{equation}
N_i^{\rm prim} = V (2J_i + 1) \int 
\frac{d^3 p}{(2\pi)^3} \, dm_i \,
\left[ \gamma_s^{-S_i}
\mbox{e}^{\frac{E_i - \mu^\alpha Q_i^\alpha}{T}} \pm 1 \right]^{-1}
\mbox{BW} (m_i),
\end{equation}
where we include phenomenologically a strangeness 
saturation factor $\gamma_s$ (with $S_i$ the total number of strange
quarks in hadron species $i$) to account for 
incomplete equilibration in this sector, 
$E_i = \sqrt{\vec p^{\, 2} + m_i^2}$, and
$\mbox{BW}$ is the Breit-Wigner distribution
(to be replaced by a $\delta$-function for stable hadrons). 
The final particle numbers are
$N_i = N_i^{\rm prim} + \sum_j \mbox{Br}^{j \to i} N_j^{\rm prim}$
due to decays of unstable particles with branching ratios 
$\mbox{Br}^{j \to i}$.  
Such a description can certainly be justified for multiplicities
measured over the whole phase-space, since many dynamical effects 
cancel out in ratios of hadron yields.

We have analyzed two data sets:
(i) NA49 $4\pi$ multiplicities of
$\langle \pi \rangle = \frac12 (\pi^+ + \pi^-)$, 
$K^\pm$, $\bar p$, $\phi$, and $N_{\rm part}$ 
(taken as the sum over all baryons)
in 6 centrality bins in the reaction Pb(158 AGeV) + Pb 
\cite{Sikler,Blume}
(it should be emphasized that protons are not included
in our analysis \cite{Spencer} since there may be a spectator 
component in non-central collisions), and
(ii) PHENIX mid-rapidity densities of $\pi^\pm$, $K^\pm$, and $p^\pm$
in the reaction Au + Au at $\sqrt{s} = 130$ AGeV in 5 centrality bins
\cite{PHENIX}. In \cite{PHENIX} it is estimated that the probability 
for reconstructing protons from $\Lambda$ decays as prompt protons is 32\% 
at $p_T$ = 1 GeV/c. This is the probability for feed-down used in 
our present fits to the mid-rapidity PHENIX data. 
The results of our fits are displayed in Fig.~1.
Specific features of the $\phi$ mesons at SPS will be discussed
elsewhere \cite{s_saturation}. 

A comparison of the individual thermal parameters of both data sets
is displayed in Fig.~2, which corrects Fig.~2 in
\cite{hirschegg_02}
(the correction of a coding error in \cite{hirschegg_02} 
lowers the temperature and brings it in 
agreement with \cite{Becattini}).
Most remarkable is the drop
of the baryo-chemical potential $\mu_B$ and the rise
of the strangeness saturation factor $\gamma_s$
when going from $\sqrt{s} = 17$ AGeV to 130 AGeV.
It would appear that $T$ remains essentially constant for  
$\sqrt{s} = 17$ AGeV, 
while at 130 AGeV it rises with $N_{\rm part}$. The parameter 
$\mu_B$ is fairly independent of the centrality, while
the strangeness saturation factor increases with centrality.
Despite the rather limited set of analyzed hadron species,
the extracted thermal parameters describe other hadron yields,
which are at our disposal in central collisions, fairly well. 
\begin{figure}[htb]
~\center
\epsfig{file=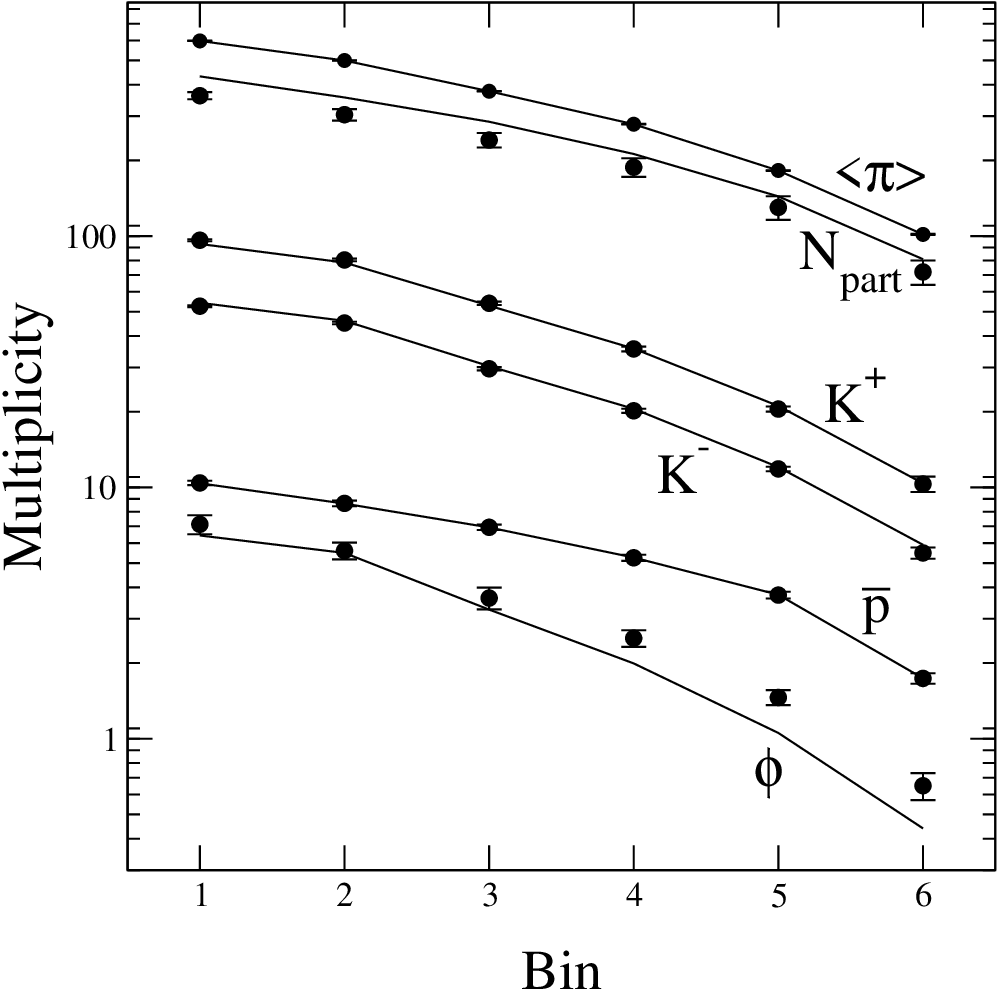,width=5cm,angle=0}
\hspace*{6mm}
\epsfig{file=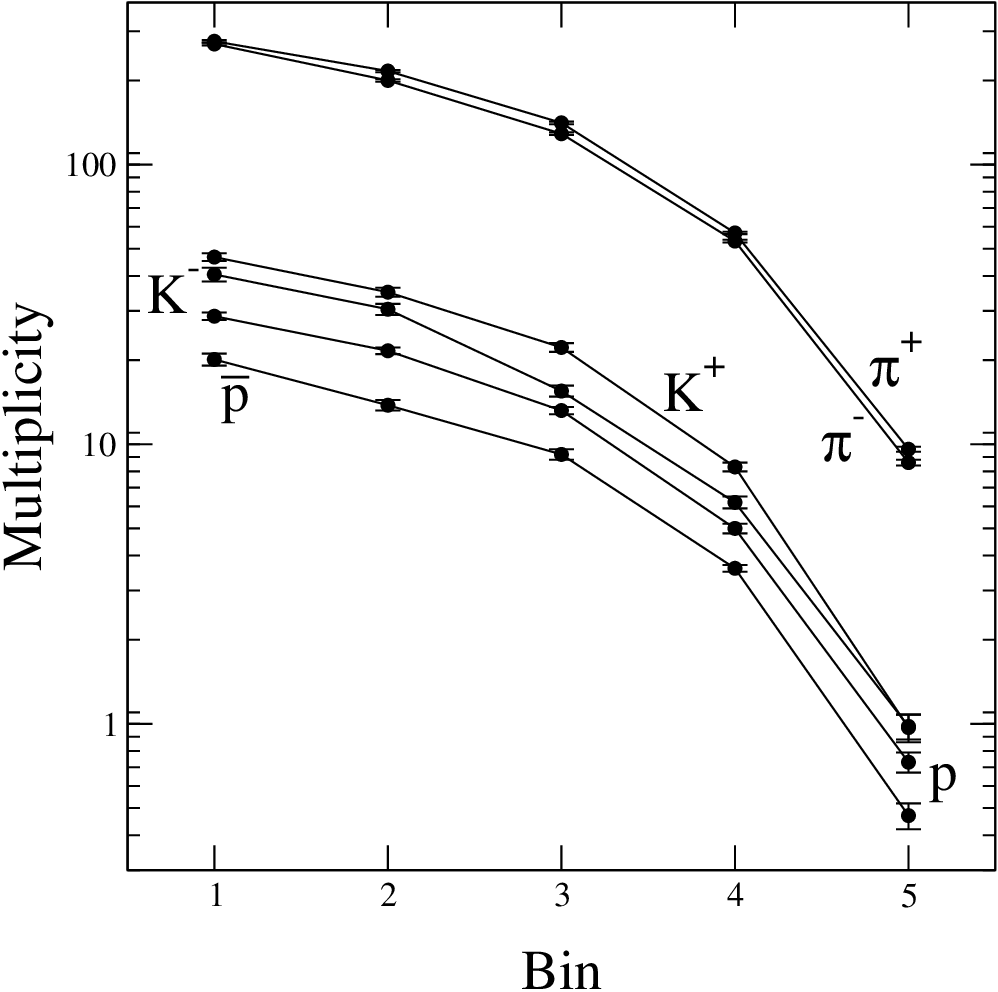,width=5cm,angle=0}

\vspace*{3mm}

\caption[]{Comparison of NA49 data (left panel, symbols, \cite{Sikler,Blume})
and PHENIX data (right panel, symbols, \cite{PHENIX})
with our model (lines).
\label{fig1}}

\vspace*{9mm}

\epsfig{file=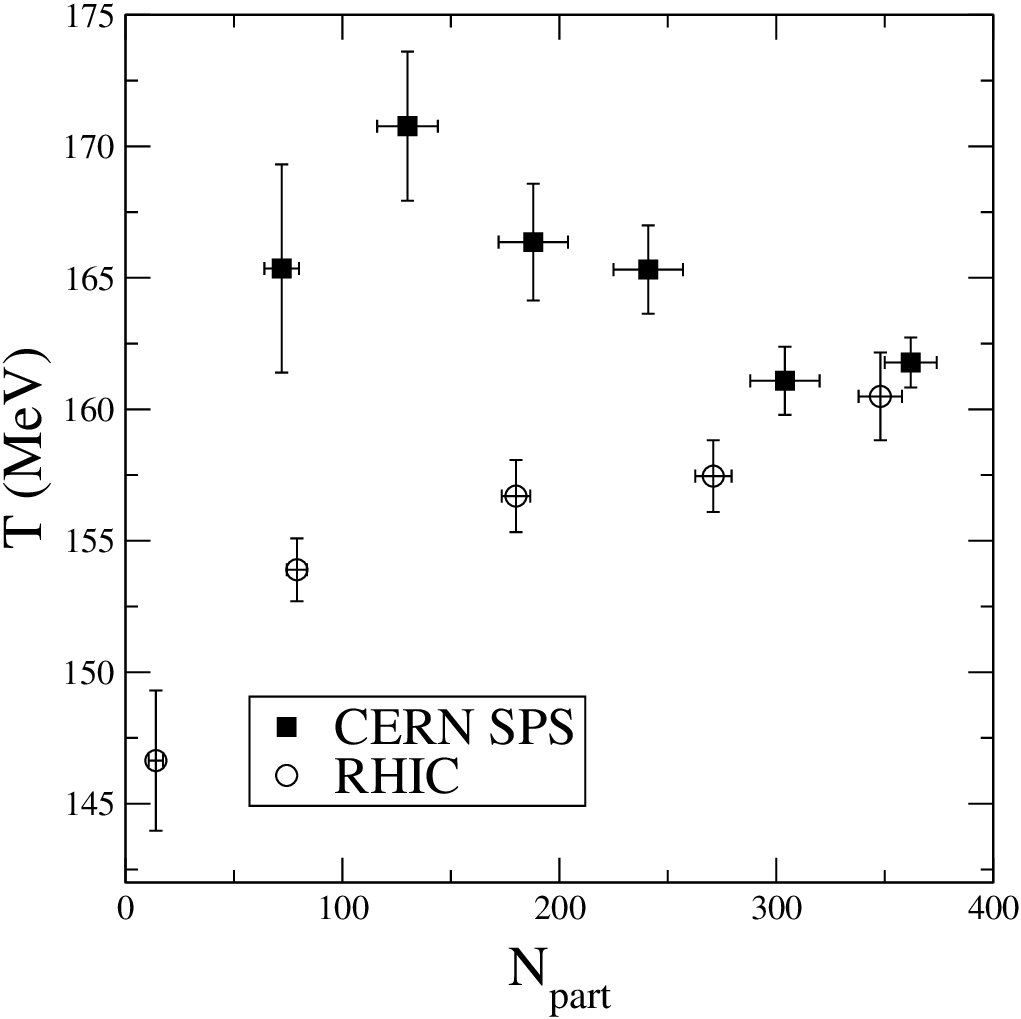,width=5cm,angle=0}
\hfill
\epsfig{file=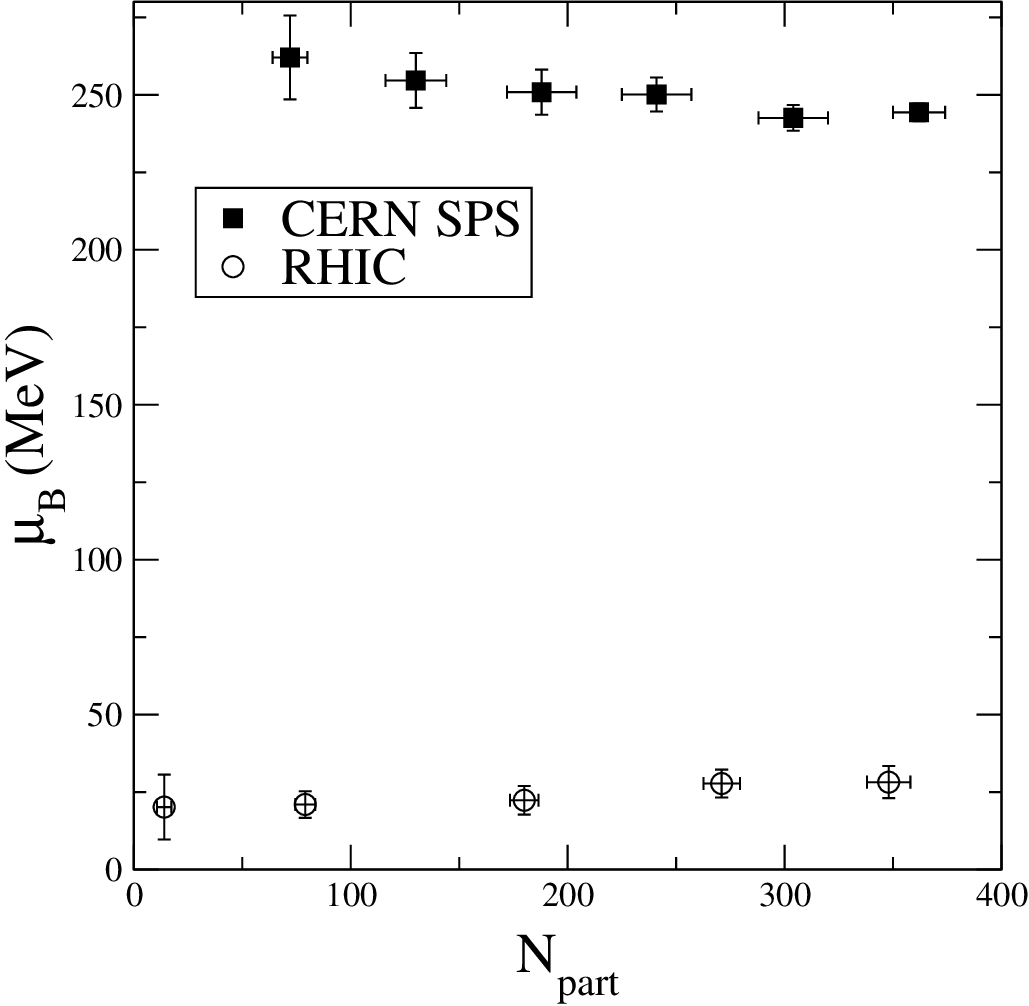,width=5cm,angle=0}
\hfill
\epsfig{file=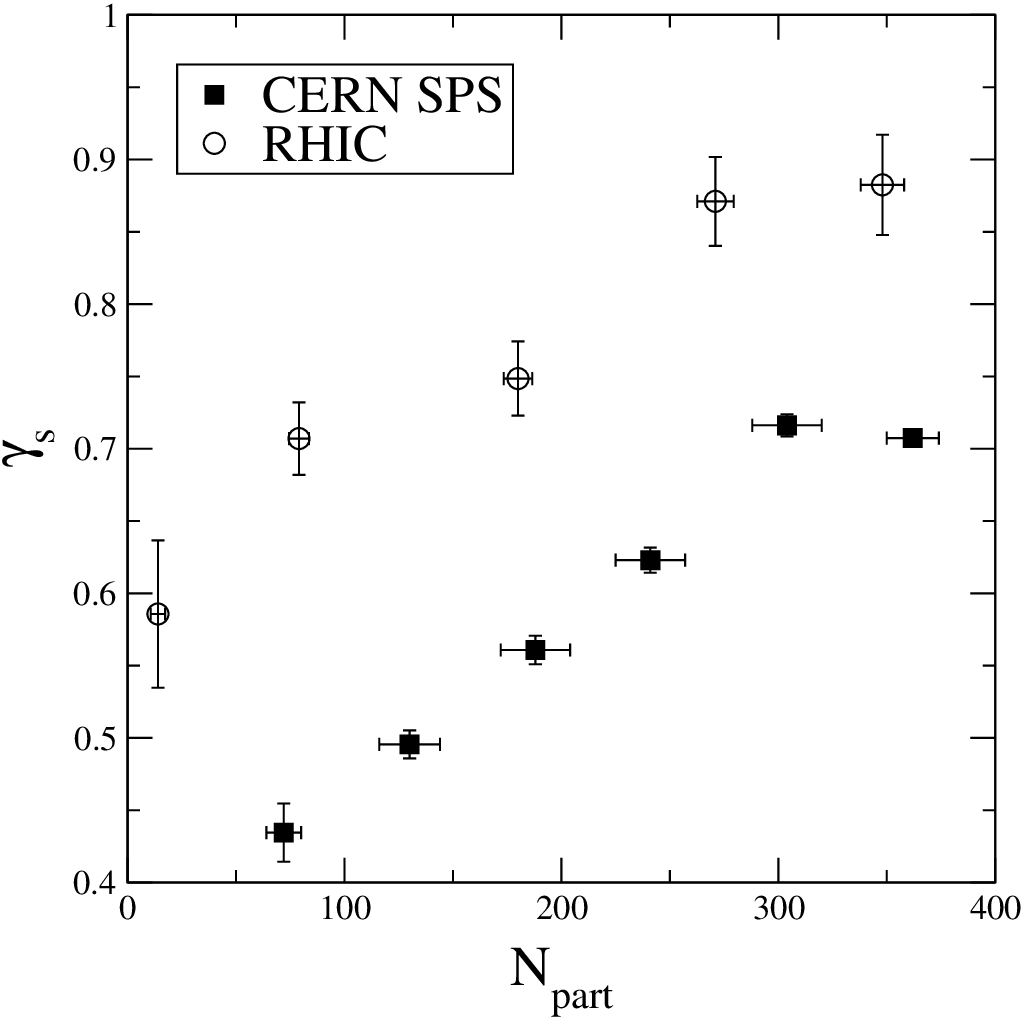,width=5cm,angle=0}

\vspace*{3mm}

\caption[]{
The temperature, baryo-chemical potential and strangeness saturation factor
as a function of $N_{\rm part}$.
Solid squares (open circles) are for the NA49 (PHENIX) 
data \cite{Sikler,Blume} (\cite{PHENIX}).
\label{fig2}}
\end{figure} 

\section{Intermediate-Mass Dileptons} 

As pointed out in \cite{Gale}, the dilepton spectra can be analyzed in the
same spirit as the hadron multiplicities and hadron momentum spectra, 
i.e., one discards any detail of the dynamics and asks only for a simple
parameterization. As a result, one gets 
for the thermal dilepton spectrum
\cite{Gale}
\begin{equation}
\frac{dN}{d^4 Q} = 
\frac{5 \alpha^2}{36 \pi^4} N_{\rm dil}
\exp \left\{ - 
\frac{M_\perp \cosh (Y - Y_{\rm cms} )}{T_{\rm dil}} \right\},
\label{eq2}
\end{equation}
\newpage 
\noindent
where $Q$ is the lepton pair's four-momentum,
$M_\perp$ its transverse mass and $Y$ its rapidity;
$Y_{\rm cms}$ denotes the fireball rapidity, and $ N_{\rm dil}$ is a
normalization factor characterizing the space-time volume of the 
fireball;
flow effects are negligible for invariant mass spectra. 
Eq.~(\ref{eq2}) is based on the quark-hadron duality
\cite{Rapp_Wambach}.

In \cite{Gale,Gallmeister} we have shown that 
the space-time averaged temperature parameter 
$T_{\rm dil} \approx 170$ MeV
(i.e., a value coinciding with the chemical freeze-out temperature)
provides a \underline{common} description of the low-mass CERES data
\cite{CERES} and the
\begin{figure}[hb]
~\center
\epsfig{file=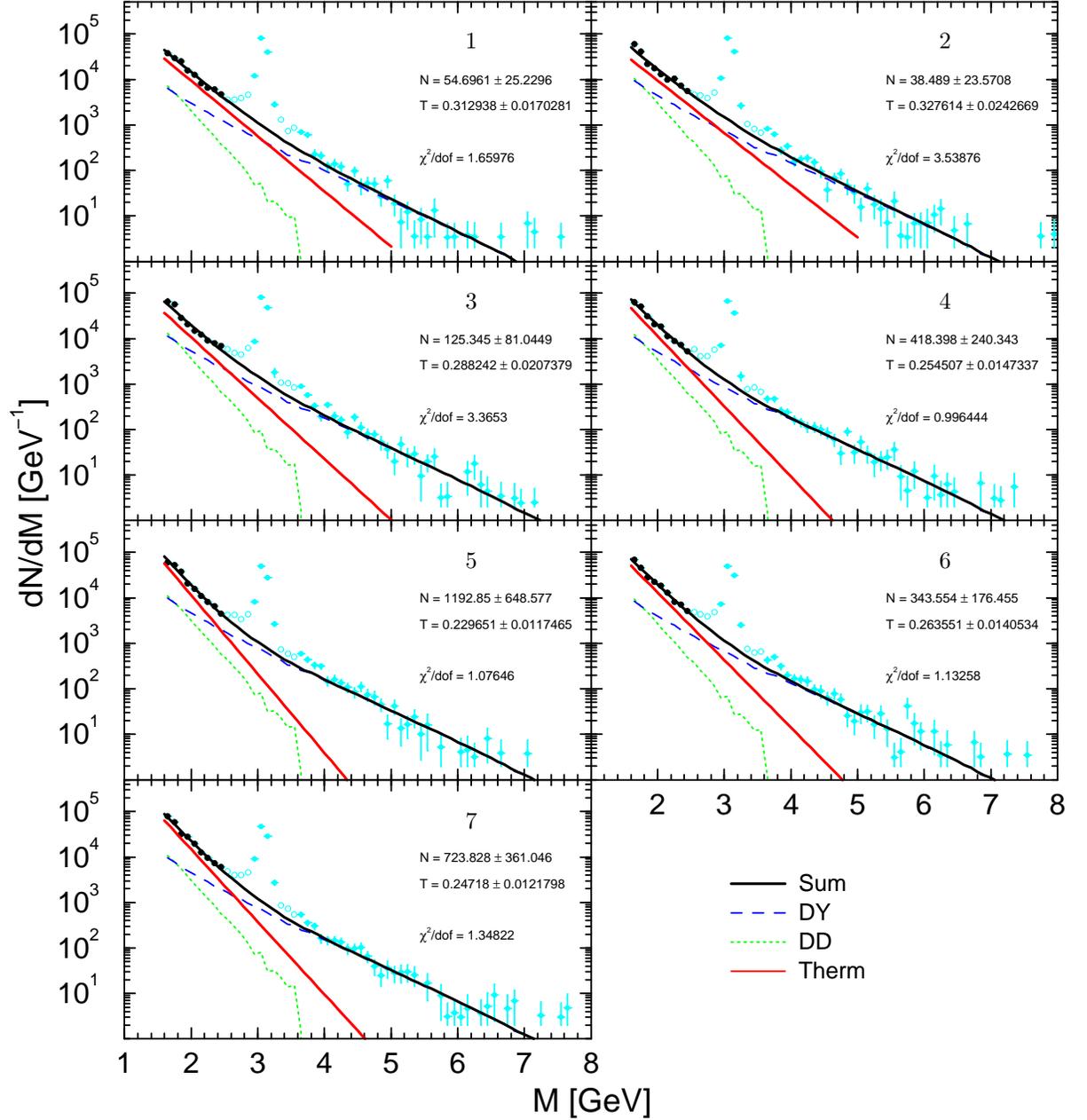,width=17cm,angle=-90}
\rput(-9.,-0.6){1}
\rput(-9.,-4.5){3}
\rput(-9.,-8.4){5}
\rput(-9.,-12.3){7}
\rput(-1.9,-0.6){2}
\rput(-1.9,-4.5){4}
\rput(-1.9,-8.4){6}

\vspace*{3mm}

\caption[]{
Fits of the thermal contribution according to Eq.~(2)
with free temperature parameter and normalization factor 
to the NA50 data \cite{Capelli}.
The data are for 7 centrality bins 
($E_T = 19, 36, 52, 67, 80, 93, 110$ GeV for bin 1 - 7).
The Drell-Yan yield and the open charm contributions are calculated
as described in detail in \cite{Gale,Gallmeister}.
\label{fig3}}

\vspace*{-2cm}

\end{figure} 
\newpage
\noindent
intermediate-mass NA50 data \cite{NA50} at $\sqrt{s} = 17$ AGeV;
also the WA98 photon data
\cite{WA98} are consistent with this value.
It was, therefore, a surprise to us that our analysis of the efficiency 
corrected and centrality binned NA50 data \cite{Capelli}
\underline{separately}
with $N_{\rm dil}$ and $T_{\rm dil}$ as free parameters
give as optimum fit a temperature scale in the order of 
250 MeV \cite{hirschegg_02}, see Figs.~3 and 4.\footnote{Note
that in Figs.~3 and 4 the used normalization of both the Drell-Yan
yield and the open charm contribution differs by a factor 1/1.5
from the subsequent figures. This covers the uncertainty in fitting
the large-$M$ Drell-Yan tail and the relative normalization
of the charm contribution.}
At this temperature the spectral shapes 
of thermal dileptons (Therm) and dileptons 
from semileptonic correlated decays of open charm mesons (DD)
are nearly identical within the NA50 acceptance.
\begin{figure}[ht]
~\center
\epsfig{file=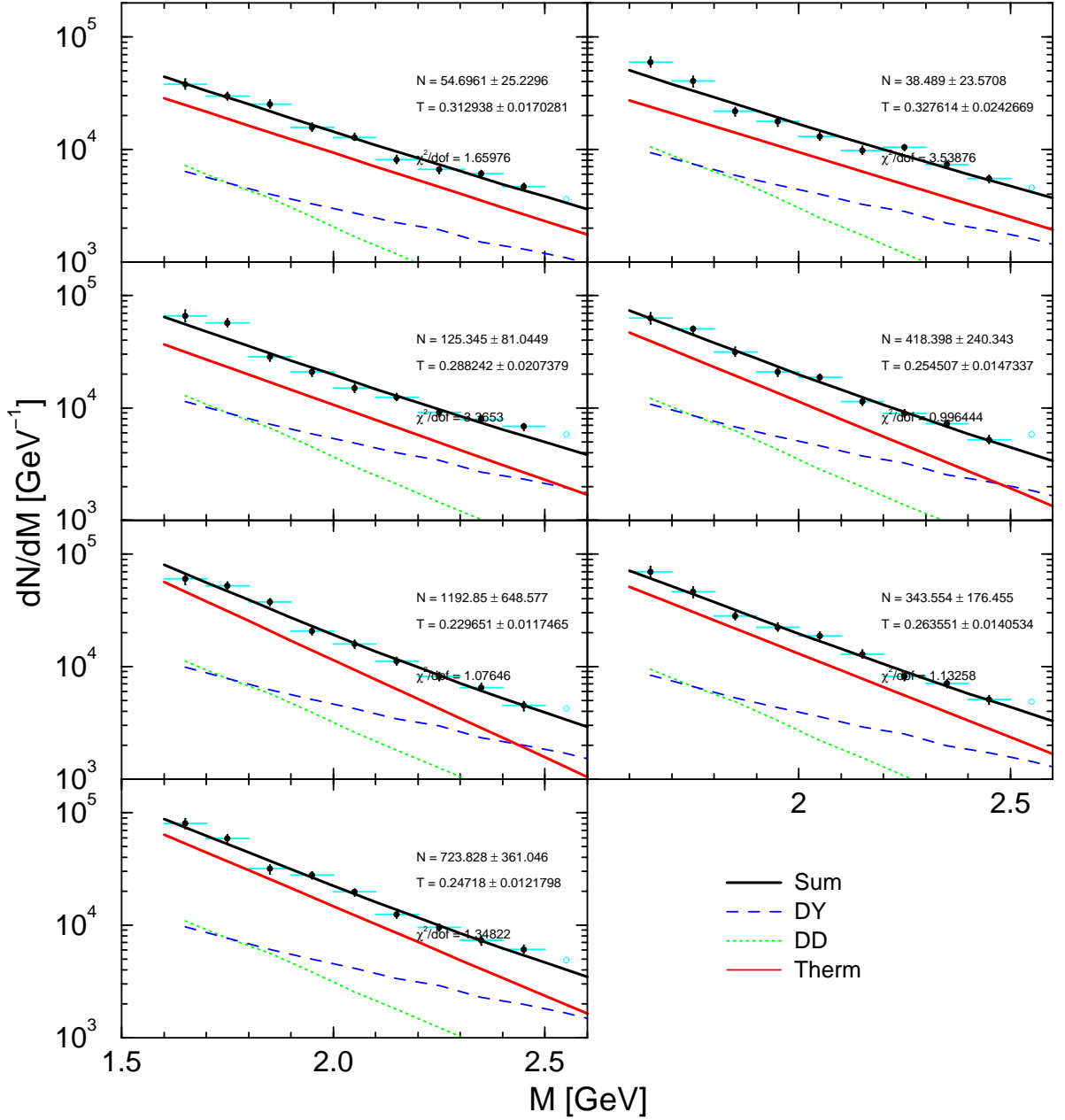,width=17cm,angle=-90}

\vspace*{3mm}

\caption[]{
As Fig.~3 but exhibiting only the intermediate-mass interval below
$J/\psi$.
\label{fig4}}

\vspace*{-2cm}

\end{figure} 
\newpage
To illustrate the sensitivity on the temperature parameter
we exhibit in Figs.~5 and 6 a comparison with the data
when using a fixed value of $T_{\rm dil} = 170$ MeV.
In contrast to the fits displayed in Figs.~3 and 4 the normalization
factor now systematically increases with centrality, as expected.
This could be interpreted as a hint that the complete disentangling
of temperature parameter and normalization factor is not fully 
adequate.\footnote{We thank R. Rapp for pointing out this possibility.}
Rather, due to the space-time evolution both quantities are correlated.
Because of uncertainties in the normalization and the shape of the
Drell-Yan (DY) and charm contributions the previously found
value of $T_{\rm dil} \approx 170$ MeV \cite{Gale,Gallmeister}
can not be excluded. 

\begin{figure}[ht]
~\center
\epsfig{file=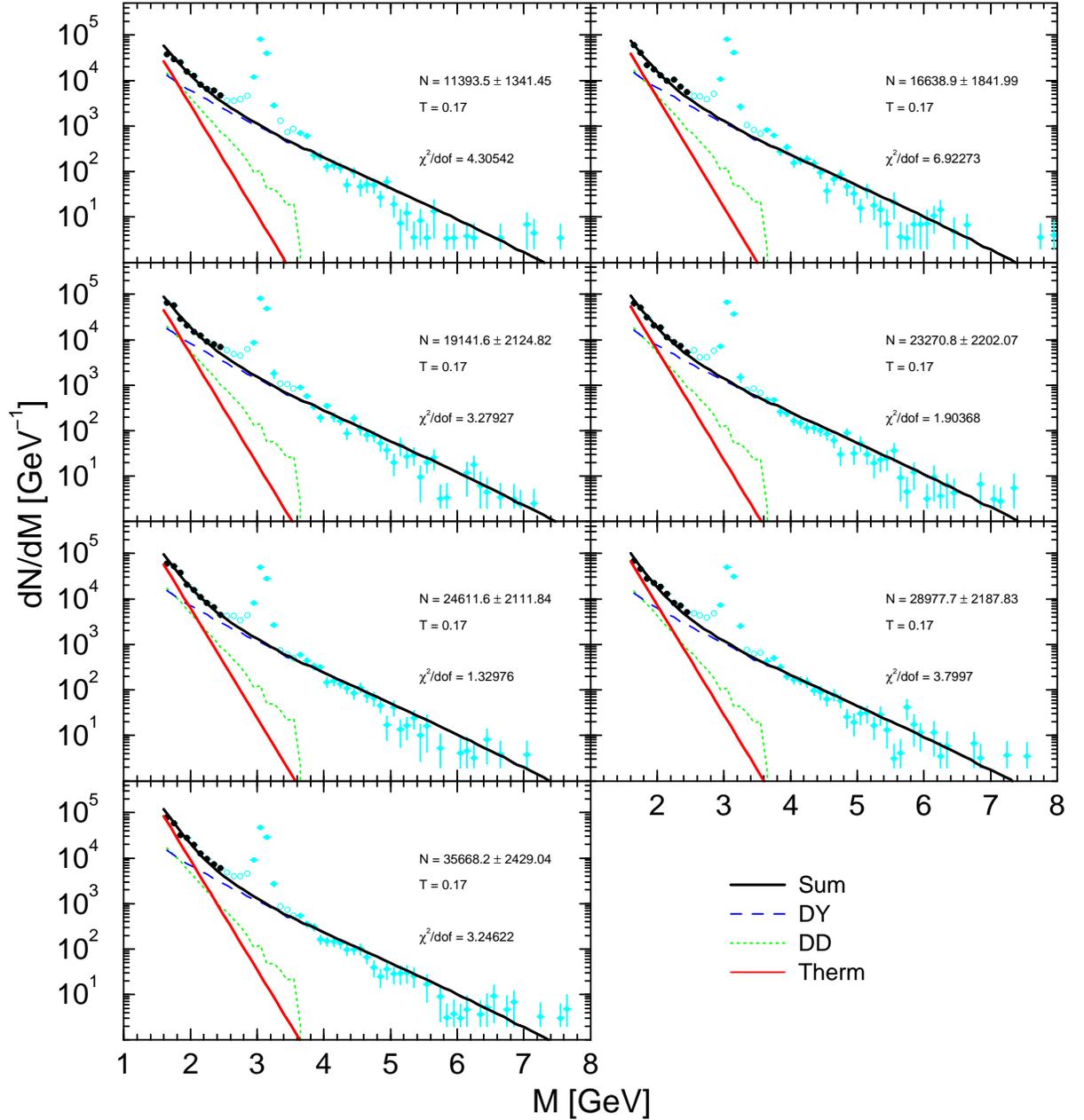,width=17cm,angle=-90}

\vspace*{3mm}

\caption[]{
As in Fig.~\ref{fig3} but for fixed $T = 170$ MeV
and free parameter $N_{\rm dil}$.
\label{fig5}}

\vspace*{-2cm}

\end{figure} 
\newpage
\begin{figure}[ht]
~\center
\epsfig{file=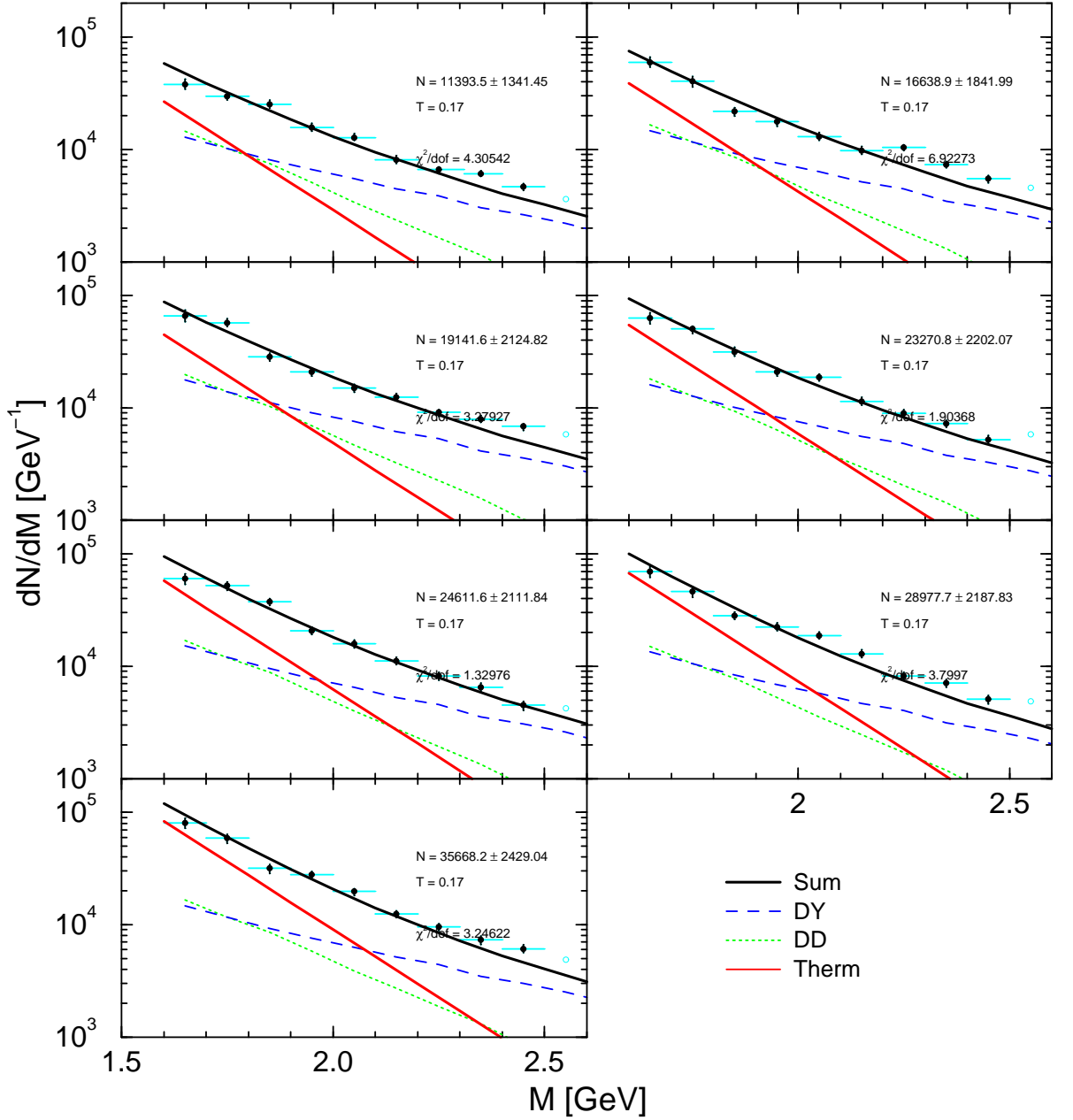,width=17cm,angle=-90}

\vspace*{3mm}

\caption[]{
As in Fig.~\ref{fig4} but for fixed $T_{\rm dil} = 170$ MeV
and free parameter $N_{\rm dil}$.
\label{fig6}}
\end{figure} 
In addition to the invariant mass spectra also transverse momentum
spectra integrated over the intermediate-mass region
$M = 1.6 \cdots 2.5$ GeV are at our disposal. At first glance,
they do not better constrain the temperature parameter $T_{\rm dil}$
since the large-$Q_\perp$ region is dominated by the
Drell-Yan yield, see Figs.~7 and 8.
The low-$Q_\perp$ region, however, favors a smaller value
of $T_{\rm dil}$, while the large-$Q_\perp$ region could be fine tuned
by a slight change of the intrinsic transverse parton momentum
(we use here $\sqrt{\langle k_\perp^2 \rangle} = 0.8$ GeV \cite{Gale}).

In contrast to the invariant mass spectra,
where the thermal yield at $T_{\rm dil} \approx 250$ MeV
has the same shape as the charm yield, the transverse mass
spectra of charm and thermal contributions look similar
at $T_{\rm dil} \approx 170$ MeV.
\newpage 
\begin{figure}[ht]
~\center
\epsfig{file=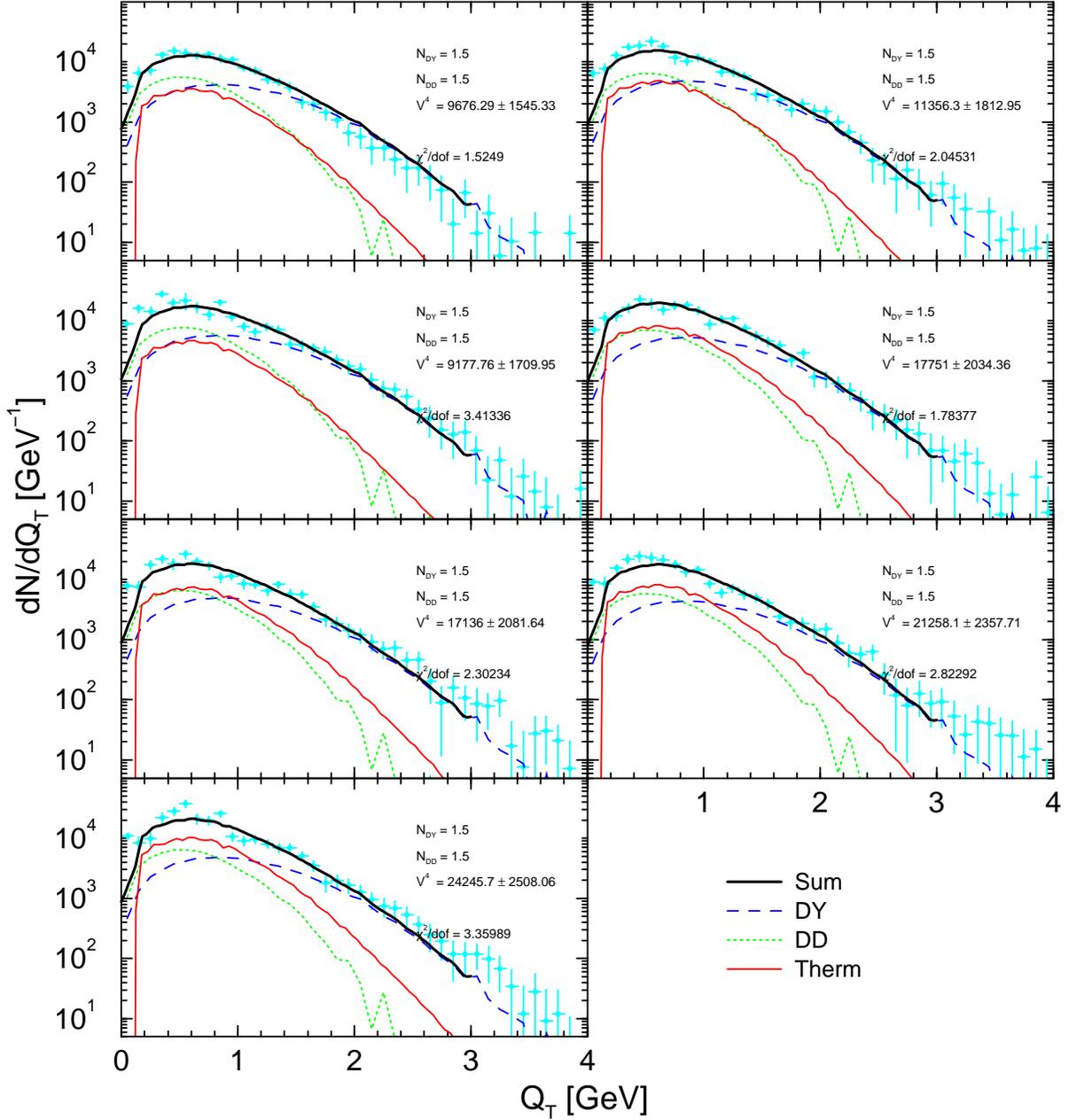,width=17cm,angle=-90}

\vspace*{3mm}

\caption[]{
Transverse momentum spectra for $M$ integrated from
1.6 $\cdots$ 2.5 GeV \protect\cite{Capelli} and 
for $T_{\rm dil} = 170$ MeV.
\label{fig7}}
\end{figure}

\section{Summary}
The analysis of hadron multiplicities indicates a
centrality independence of the chemical
freeze-out temperature and baryo-chemical potential for 
$\sqrt{s} = 17$ AGeV, while at $\sqrt{s} = 130$ AGeV
the baryo-chemical potential is constant but the temperature 
increases with centrality.
The strangeness saturation increases with centrality
for both $\sqrt{s} = 17$ AGeV and 130 AGeV and appears to approach
unity with further increasing energy and maximum centrality. 
\newpage
\begin{figure}[ht]
~\center
\epsfig{file=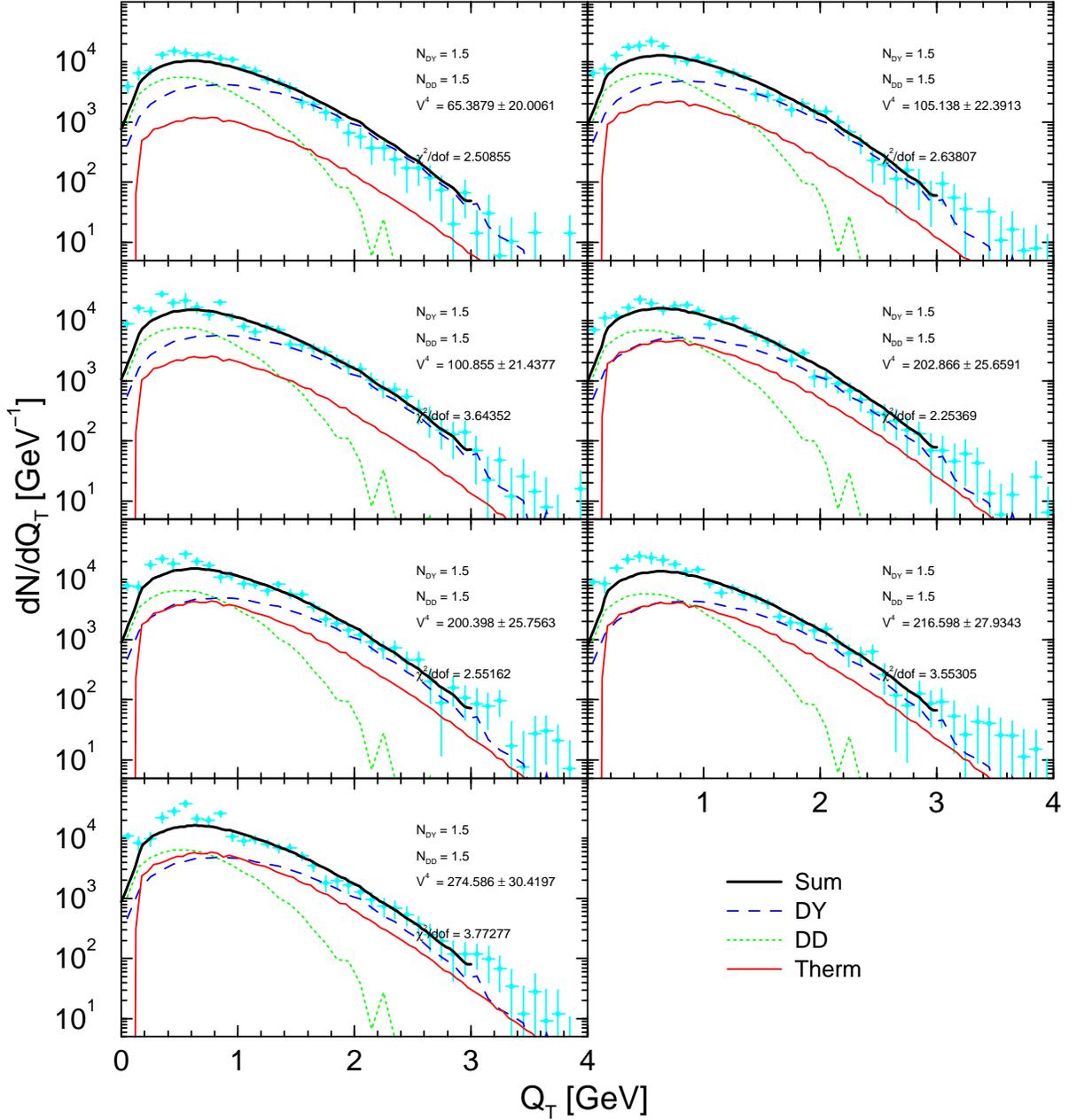,width=17cm,angle=-90}

\vspace*{3mm}

\caption[]{
As Fig.~7 but for $T_{\rm dil} = 250$ MeV.
$\chi^2 / d.o.f.$ is marginally larger than for $T_{\rm dil} = 170$
MeV
in Fig.~7.
\label{fig8}}

\vspace*{-1mm}

\end{figure} 

The analysis of the intermediate-mass dilepton spectra
taken by the NA50 collaboration 
is not yet conclusive. Our parameterization of the thermal
contribution is based on quark-hadron duality where the shape and
the normalization are not interrelated. We find an
indication that the space-time averaged temperature determining 
the shape of dilepton spectra stays fairly independent of centrality,
but is significantly larger than
previously anticipated in \cite{Gale,Gallmeister} thus being
in agreement with values deduced in \cite{dinesh} from photon data.
It should be noted, however, that the analysis is sensitive to details of
the Drell-Yan and open charm dilepton yields, both their shape and
normalization. In view of these inherent uncertainties, the
centrality independent value of $T_{\rm dil} \approx 170$ MeV
appears still compatible with the NA50 data. 

\section*{Acknowledgment(s)}
We are grateful to M. Gazdzicki, 
F. Sikler, Ch. Blume, and C. Lourenco for discussions on the
experimental data and K. Redlich for continuous discussions
concerning the thermal model. 
The work is supported by BMBF 06DR921.


\begin{thebibliography}{99}  
\bibitem{hirschegg_02} B. K\"ampfer, J. Cleymans, K. Gallmeister,
S. Wheaton, hep-ph/0202134,
Proc.\ of the International Workshop XXX
on Gross Properties of Nuclei and Nuclear Excitations,
Hirschegg, Austria, January 13 - 19, 2002,
(Eds.) M. Buballa, W. N\"orenberg, B.-J. Schaefer, J. Wambach,
p. 158
\bibitem{abundance} 
P.~Braun-Munzinger et al., 
Phys.~Lett.~{\bf B344} (1995) 43, 
{\bf B365} (1996) 1, 
{\bf B465} (1999) 15, 
{\bf B518} (2001) 415\\
J.~Cleymans, K.~Redlich, Phys.~Rev.~Lett.~{\bf 81} (1998) 5284
\bibitem{Becattini} F.~Becattini, J.~Cleymans, A.~Keranen, E.~Suhonen,
K.~Redlich,\\
Phys.~Rev. ~{\bf C64} (2001) 024901 
\bibitem{Sikler} F.~Sikler (NA49 collaboration), 
Nucl.~Phys.~{\bf A661} (1999) 45c
\bibitem{Blume} V.~Friese (NA49 collaboration), 
Nucl.~Phys.~{\bf A698} (2002) 487c 
\bibitem{Spencer} J.~Cleymans, B.~K\"ampfer, S.~Wheaton, 
Phys.~Rev.~{\bf C65} (2002) 027901
\bibitem{PHENIX} K.~Adcox et al. (PHENIX collaboration), 
nucl-ex/0112006
\bibitem{s_saturation} J. Cleymans, B. K\"ampfer, S.M. Wheaton,
in preparation
\bibitem{Gale} K.~Gallmeister, B.~K\"ampfer, O.P.~Pavlenko, C.~Gale,
Nucl.~Phys.~{\bf A688} (2001) 939, {\bf A698} (2002) 424c
\bibitem{Rapp_Wambach} R. Rapp, J. Wambach, Adv. Nucl. Phys. {\bf 25}
(2000) 1   
\bibitem{Gallmeister} K.~Gallmeister, B.~K\"ampfer, O.P.~Pavlenko,
Phys.~Lett.~{\bf B473} (2000) 20, 
Phys.~Rev.~{\bf C62} (2000) 057901 
\bibitem{CERES} B. Lenkeit (CERES collaboration), 
Nucl. Phys. {\bf A661} (1999) 23c 
\bibitem{NA50} E. Scomparin ( NA50 collaboration),
Nucl. Phys. {\bf A610} (1996) 331, J. Phys. {\bf G25} (1999) 235c
\bibitem{WA98} M.M. Aggarwal et al. (WA98 collaboration),
Phys. Rev. Lett. {\bf 85} (2000) 3595 
\bibitem{Capelli} L.~Capelli (NA50 collaboration), Ph.\ D.\ thesis,
University of Lyon, 2000,
Nucl. Phys. {\bf A698} (2002) 539c
\bibitem{dinesh} D.K. Srivastava, B. Sinha, I. Kvasnikova, C. Gale,
Nucl. Phys. {\bf A698} (2002) 432c \\
D.K. Srivastava, B. Sinha, Phys. Rev. {\bf C64} (2001) 034902  
\end{thebibliography}
\end{document}